\documentclass[aps,pra,twocolumn,groupedaddress,showpacs]{revtex4}
\usepackage{amssymb}
\usepackage{amsmath}
\usepackage{dcolumn}
\usepackage{bm}
\usepackage{graphicx}
\usepackage{mathrsfs}

\begin{document}

\title{Two-photon superbunching of pseudothermal light in a Hanbury Brown-Twiss interferometer}
\author{Bin Bai $^{1}$}
\author{Jianbin Liu $^{1}$}
\email{liujianbin@xjtu.edu.cn}
\author{Yu Zhou $^{2}$}
\author{Huaibin Zheng $^{1}$}
\author{Hui Chen $^{1}$}
\author{Songlin Zhang $^{1}$}
\author{Yuchen He $^{1}$}
\author{Fuli Li $^{2}$}
\author{Zhuo Xu $^{1}$}

\affiliation{$^1$Electronic Materials Research Laboratory, Key Laboratory of the Ministry of Education \& International Center for Dielectric Research, Xi'an Jiaotong University, Xi'an 710049, China}
\affiliation{$^2$MOE Key Laboratory for Nonequilibrium Synthesis and Modulation of Condensed Matter, and Department of Applied Physics, Xi'an Jiaotong University, Xi'an 710049, China}

\begin{abstract}
Two-photon superbunching of pseudothermal light is observed with single-mode continuous-wave laser light in a linear optical system. By adding more two-photon paths via three rotating ground glasses, $g^{(2)}(0)=7.10\pm0.07$ is experimentally observed. The second-order temporal coherence function of superbunching pseudothermal light is theoretically and experimentally studied in detail. It is predicted that the degree of coherence of light can be increased dramatically by adding more multi-photon paths. For instance, the degree of the second- and third-order coherence of the superbunching pseudothermal light with five rotating ground glasses can reach 32 and 7776, respectively. The results are helpful to understand the physics of superbunching and to improve the visibility of thermal light ghost imaging.
\end{abstract}

\maketitle

\section{introduction}

Two-photon bunching was first observed by Hanbury Brown and Twiss in 1956, in which randomly emitted photons in thermal light are not random \cite{hbt,hbt2}.  In their experiments, they found that photons in thermal light have the tendency to come in bunches rather than randomly when the two detectors are in the symmetrical positions \cite{hbt3}. Their surprising results drew lots of attentions at that time. Some physicists repeated Hanbury Brown and Twiss's experiments and got negative results \cite{adam,brannen}. It was later understood that the reason for the negative results in the experiments \cite{adam,brannen} is due to that the response time of the detection system is longer than the coherence time of the light field \cite{mandel-book}. Both classical theory \cite{hbt3} and quantum theory \cite{glauber1,glauber2} were employed to interpret Hanbury Brown and Twiss's experimental results. It is now well-known that quantum and classical theories are equivalent in interpreting two-photon bunching of thermal light \cite{glauber2,sudarshan}. However, the quantum interpretation of two-photon bunching greatly advance the development of quantum optics \cite{glauber1,glauber2}. Hanbury Brown and Twiss's experiments \cite{hbt,hbt2}, together with Glabuer's quantum optical coherence theory, are usually regarded as the cornerstones of modern quantum optics \cite{glauber3}.

Photon superbunching is employed to describe the properties of light in which photons are more bunched than the ones in thermal light. Mathematically, $N$-photon superbunching is defined as the degree of $N$th-order coherence of light is larger than the one of thermal light  \cite{loudon-book,ficek-book}, where $N$ is an integer greater than 2. For instance, the degree of second-order coherence of thermal light equals 2 \cite{hbt,hbt2} and there is two-photon superbunching if the degree of second-order coherence is greater than 2. The degree of $N$th-order coherence of thermal light equals $N!$ \cite{liu-2009}. There is $N$-photon superbunching when the degree of  $N$th-order coherence of light is greater than $N!$.

Two-photon superbunching has been studied extensively in quantum optics and quantum information \cite{mandel-book,loudon-book,chuang-book}. It is usually generated by nonlinear interaction between light and matter \cite{lipeles,kaul,kocher,akhmanov,swain,auffeves,hoi,grujic,bhatti,albert,jahnke,redlich,klyshko-1970,burnham,alley,ou}. Two-photon superbunching was first observed in a single three-level atom system pumped by a light beam\cite{lipeles,kaul,kocher,akhmanov,swain}. The efficiency of collecting photons is very low due to the generated photon pairs are incident to all $4\pi$ directions. Later, it was found that the collecting efficiency could be increased by putting atoms  \cite{auffeves,hoi,grujic,bhatti} or quantum dots \cite{albert,jahnke,redlich} into a cavity. Nowadays, the most common method to generate two-photon superbunching in laboratory is spontaneous parametric down conversion in a nonlinear crystal pumped by laser light  \cite{klyshko-1970,burnham,alley,ou}. However, the efficiency of generating superbunching photon pairs via nonlinear interaction is extremely low, which is below $10^{-12}$ \cite{boyd-book}. Besides low efficiency, the experimental setup of generating two-photon superbunching by nonlinear interaction usually needs careful alignment \cite{lipeles,kaul,kocher,akhmanov,swain,auffeves,hoi,grujic,bhatti,albert,jahnke,redlich,klyshko-1970,burnham,alley,ou}. If linear system can be employed to generate two-photon superbunching, it will be convenient to study the second-order coherence of light. Recently, we have observed two-photon spatial and temporal superbunching by employing classical light in linear optical systems \cite{hong,zhou-2017}. By employing a pinhole and two rotating ground glasses, we observed the degree of second-order coherence of light equals $3.66 \pm 0.02$ in \cite{zhou-2017}. It is predicted that the system can be generalized to reach larger value of  degree of second-order coherence \cite{zhou-2017}. In this paper, we will study in detail how the second-order coherence function is influenced by the system and further increase the degree of second-order coherence. The results are helpful to understand the physics of superbunching and to improve the visibility of thermal light ghost imaging \cite{shih-book}.

The paper is organized as follows. Employing two-photon interference theory to interpret two-photon superbunching of pseudothermal light can be found in Sec. \ref{theory}. Experimental study of two-photon superbunching effect of pseudothermal light is in Sec. \ref{experiment}. The discussions and conclusions are in Sec. \ref{discussion} and \ref{conclusion}, respectively.

\section{Theory}\label{theory}

Both quantum and classical theories can be employed to interpret two-photon superbunching effect of supebunching pseudothermal light \cite{hbt3,glauber1,glauber2,sudarshan}. In our earlier studies, we have employed two-photon interference based on the superposition principle in Feynman's path integral theory to interpret the second-order interference of light \cite{zhou-2017,liu-2010,liu-2013,liu-2014,liu-2015,liu-2016,liu-2017}, which is proved to be helpful to understand the connection between  the physical interpretations and mathematical calculations. In this section, the same method will be employed to calculate the second-order coherence function of superbunching pseudothermal light.

\begin{figure}[htb]
    \centering
    \includegraphics[width=50mm]{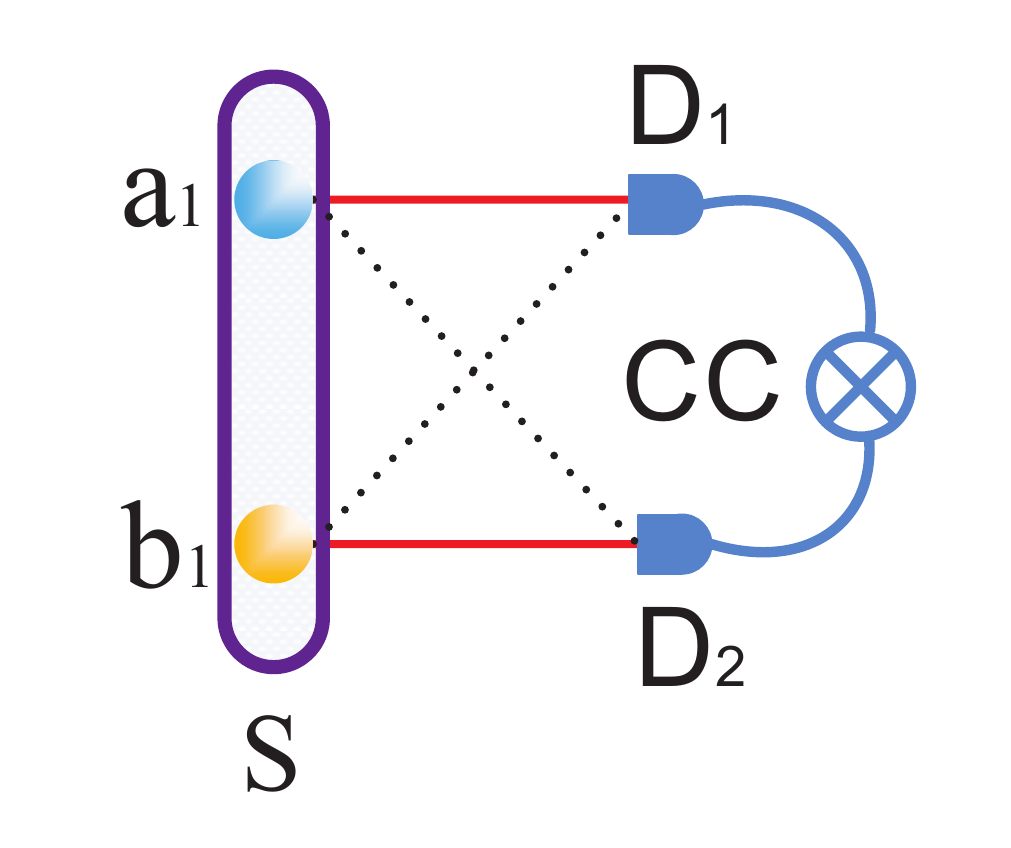}
    \caption{Two-photon interference of thermal light in a HBT interferometer. S is a thermal light source. a$_1$ and b$_1$ are two possible positions for two photons, respectively. D$_1$ and D$_2$ are two single-photon detectors. CC is two-photon coincidence count detection system. The red solid lines and black dot lines correspond to two different ways to trigger a two-photon coincidence count.} \label{1}
\end{figure}

Two-photon interference theory had been employed by many physicists to interpret two-photon bunching of thermal light. For instance, it was first pointed out by Fano that there are two different alternatives for two photons in thermal light to trigger a two-photon coincidence count in a Hanbury Brown-Twiss (HBT) interferometer \cite{fano}. As shown in Fig. \ref{1}, the first alternative is photon a$_1$ (short for photon at position a$_1$) is detected by D$_1$ (short for detector 1) and photon b$_1$ is detected by D$_2$. The second alternative is  photon a$_1$ is detected by D$_2$ and photon b$_1$ is detected by D$_1$. If these two different alternatives are indistinguishable, the second-order coherence function is
\begin{equation}\label{g2-1}
G^{(2)}(\vec{r}_{1},t_{1};\vec{r}_{2},t_{2})=\langle |A_{a1D1}A_{b1D2}+A_{a1D2}A_{b1D1}|^2 \rangle,
\end{equation}
where $(\vec{r}_{j},t_{j})$ is the space-time coordinate of the photon detection event at D$_j$ ($j=1$ and 2). $\langle...\rangle$ is ensemble average by taking all the possible realizations into account, which is equivalent to time average for a stationary and ergodic system \cite{mandel-book}. $A_{a1D1}$ is the probability amplitude for photon a$_1$  is detected by D$_1$ and the meanings of other symbols are defined similarly. $A_{a1D1}A_{b1D2}$ and $A_{a1D2}A_{b1D1}$ are the probability amplitudes corresponding to the above two alternatives to trigger a two-photon coincidence count event, respectively. The superposition of these two probability amplitudes, which is usually called two-photon interference \cite{shih-book}, is the quantum interpretation of two-photon bunching of thermal light. Feynman himself also presented similar interpretation for two-photon bunching of thermal light in one of his lectures on quantum electrodynamics \cite{feynman-qed}. Similar interpretation was also employed by other physicists to interpret the second-order coherence of thermal light \cite{scully-book,shih-2005}.

\begin{figure}[htb]
    \centering
    \includegraphics[width=80mm]{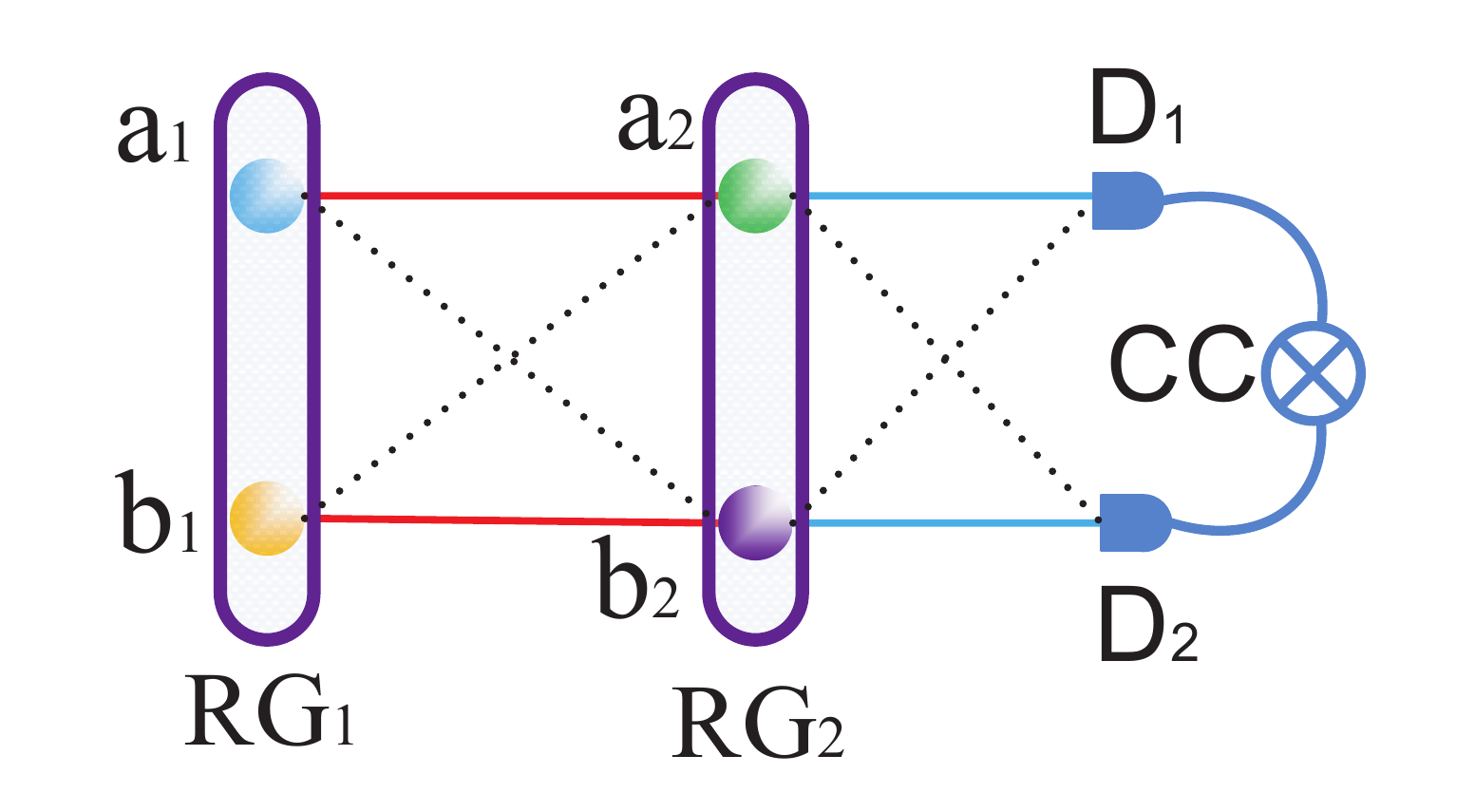}
    \caption{Two-photon interference for more than two alternatives in a HBT interferometer. S is a  thermal light source. a$_j$ and b$_j$ are two possible positions for two photons on the $j$th rotating ground glass, RG$_j$, respectively ($j=1$ and 2). D$_1$ and D$_2$ are two single-photon detectors. CC is two-photon coincidence count detection system. The combinations of solid lines and dot lines correspond to possible ways to trigger a two-photon coincidence count.} \label{2}
\end{figure}

When a single-mode continuous-wave laser light beam is incident to a rotating ground glass (RG), the scattered light after RG is usually called pseudothermal light \cite{martienssen}. There is two-photon bunching for photons in pseudothermal light and the quantum interpretation of this phenomenon is the same as the one in \cite{fano,feynman-qed,scully-book,shih-2005}. Similar interpretation is valid when there are more than two different ways to trigger a two-photon coincidence count \cite{hong,zhou-2017,liu-2010,liu-2013,liu-2014,liu-2015,liu-2016,liu-2017}. For instance, there are four different ways for photons a$_1$ and b$_1$ to trigger a two-photon coincidence count in the scheme shown in Fig. \ref{2}. RG$_1$ and RG$_2$ are two rotating ground glasses. The meanings of other symbols in Fig. \ref{2} are similar as the ones in Fig. \ref{1}. The first way is photon a$_1$ goes to a$_2$ and then is detected by D$_1$ and photon a$_2$ goes to b$_2$ and then is  detected by D$_2$, in which the probability amplitude can be written as $A_{a1a2}A_{a2D1}A_{b1b2}A_{b2D2}$. Other three different alternatives correspond to $A_{a1a2}A_{a2D2}A_{b1b2}A_{b2D1}$, $A_{a1b2}A_{b2D1}A_{b1a2}A_{a2D2}$, and $A_{a1b2}A_{b2D2}A_{b1a2}A_{a2D1}$. If all the four different alternatives are indistinguishable, the second-order coherence function in the scheme shown in Fig. \ref{2} is \cite{feynman-path,zhou-2017}
\begin{eqnarray}\label{g2-2}
&&G^{(2)}(\vec{r}_{1},t_{1};\vec{r}_{2},t_{2})\nonumber\\
&=&\langle |A_{a1a2}A_{a2D1}A_{b1b2}A_{b2D2}+A_{a1a2}A_{a2D2}A_{b1b2}\nonumber\\
&&A_{b2D1}+A_{a1b2}A_{b2D1}A_{b1a2}A_{a2D2}+A_{a1b2}A_{b2D2}\nonumber\\
&&A_{b1a2}A_{a2D1}|^2 \rangle\nonumber\\
&=& \langle |(A_{a1a2}A_{b1b2}+A_{a1b2}A_{b1a2})(A_{a2D1}A_{b2D2}\nonumber\\
&&+A_{a2D2}A_{b2D1})|^2 \rangle.
\end{eqnarray}
The last line of Eq. (\ref{g2-2}) indicates that there may exist a convenient way to present the probability amplitudes for all the different alternatives when there are more than two RGs.

The calculations above can be generalized to the case when there are $N$ RGs. It has been proved that there are $2^N$ different ways to trigger a two-photon coincidence count when there are $N$ RGs in the scheme similar as the one in Fig. \ref{2}\cite{zhou-2017}. Let us assume that all the $2^N$ probability amplitudes can be expressed as
 \begin{eqnarray}\label{g2-n}
&&(A_{a1a2}A_{b1b2}+A_{a1b2}A_{b1a2})(A_{a2a3}A_{b2b3}+A_{a2b3}\nonumber\\
&&A_{b2a3})...[A_{a(N-1)aN}A_{b(N-1)bN}+A_{a(N-1)bN}\nonumber\\
&&A_{b(N-1)aN}]\times[A_{aND1}A_{bND2}+A_{aND2}A_{bND1}]\nonumber\\
&=& \prod_{j=1}^{N-1}[A_{a(j)a(j+1)}A_{b(j)b(j+1)}+A_{a(j)b(j+1)}A_{b(j)a(j+1)}]\nonumber\\
&&\times[A_{aND1}A_{bND2}+A_{aND2}A_{bND1}]\nonumber\\
&\equiv& \Lambda_{N-1}\times[A_{aND1}A_{bND2}+A_{aND2}A_{bND1}],
\end{eqnarray}
where $\Lambda_{N-1}$ equals $\prod_{j=1}^{N-1}[A_{a(j)a(j+1)}A_{b(j)b(j+1)}+A_{a(j)b(j+1)}A_{b(j)a(j+1)}]$.  There are two more possible positions for the detected two photons if another RG is added after the $N$th RG. The first way is photon a$_{N+1}$ is detected by D$_1$ and photon b$_{N+1}$ is detected by D$_2$. The probability amplitudes can be expressed as
 \begin{eqnarray}\label{g2-n1}
&&\Lambda_{N-1}\times[A_{aNa(N+1)}A_{bNb(N+1)}+A_{aNb(N+1)}\nonumber\\
&&A_{bNa(N+1)}]
\times A_{a(N+1)D1}A_{b(N+1)D2},
\end{eqnarray}
where the first two terms are obtained by replacing $D1$ and $D2$ with $a(N+1)$ and $b(N+1)$ in Eq. (\ref{g2-n}), respectively. The second way to trigger a two-photon coincidence count by adding the $(N+1)$th RG is photon a$_{N+1}$ is detected by D$_2$ and photon b$_{N+1}$ is detected by D$_1$. With the same method above, the corresponding probability amplitudes can be written as
 \begin{eqnarray}\label{g2-n2}
&&\Lambda_{N-1}\times[A_{aNa(N+1)}A_{bNb(N+1)}+A_{aNb(N+1)}\nonumber\\
&&A_{bNa(N+1)}]\times A_{a(N+1)D2}A_{b(N+1)D1}.
\end{eqnarray}
The probability amplitudes for $(N+1)$ RGs are the sum of Eqs. (\ref{g2-n1}) and (\ref{g2-n2}). It is straightforward to have
 \begin{eqnarray}\label{g2-n3}
&&\Lambda_{N-1}\times[A_{aNa(N+1)}A_{bNb(N+1)}+A_{aNb(N+1)}\nonumber\\
&&A_{bNa(N+1)}]\times A_{a(N+1)D1}A_{b(N+1)D2}\nonumber\\
&&+\Lambda_{N-1}\times[A_{aNa(N+1)}A_{bNb(N+1)}+A_{aNb(N+1)}\nonumber\\
&&A_{bNa(N+1)}]\times A_{a(N+1)D2}A_{b(N+1)D1}\nonumber\\
&=&\Lambda_{N}\times[A_{a(N+1)D1}A_{b(N+1)D2}+A_{a(N+1)D2}\nonumber\\
&&A_{b(N+1)D1}].
\end{eqnarray}
We have generalized Eq. (\ref{g2-n}) for $N$ RGs to Eq. (\ref{g2-n3}) for $N+1$ RGs.  When $N$ equals 1 and 2 in Eq.  (\ref{g2-n}), Eqs. (\ref{g2-1}) and (\ref{g2-2}) are obtained. Hence our results are valid for all $N$ ($N$ is a positive integer) RGs in the scheme similar as the one shown in Fig. \ref{2}.

If all the $2^N$ different ways to trigger a two-photon coincidence count are indistinguishable, the second-order coherence function for $N$ RGs is \cite{feynman-qed,zhou-2017}
\begin{eqnarray}\label{g2-n4}
&&G^{(2)}(\vec{r}_{1},t_{1};\vec{r}_{2},t_{2})\nonumber\\
&=& \langle |\prod_{j=1}^{N-1}[A_{a(j)a(j+1)}A_{b(j)b(j+1)}+A_{a(j)b(j+1)}A_{b(j)a(j+1)}] \nonumber\\
&&\times [A_{aND1}A_{bND2}+A_{aND2}A_{bND1}]|^2 \rangle.
\end{eqnarray}
Substituting the detail expressions for probability amplitudes into Eq. (\ref{g2-n4}), the second-order coherence function can be calculated. Here we will concentrate on the second-order temporal coherence by assuming a$_j$ and b$_j$ ($j=1$, 2, ..., and $N$) are in the symmetrical positions and so are D$_1$ and D$_2$. It can be realized by assuming the size of light spot on every RG is a point and these two detectors are in the symmetrical positions in the HBT interferometer.

The temporal photon propagator for a point light source is \cite{qft,zhou-2017}
\begin{equation}\label{green-t}
K_{\alpha\beta}\propto e^{-i\omega_{\alpha}(t_{\beta}-t_\alpha)},
\end{equation}
which is the same as Green function in classical optics \cite{born-book}. $\omega_{\alpha}$ is the frequency of light scattered by RG$_{\alpha}$. $t_{\alpha}$ and $t_\beta$ are the time coordinates for photon at different instants. Substituting Eq. (\ref{green-t}) into Eq. (\ref{g2-n4}) and with the same method as the one in \cite{zhou-2017}, it is straightforward to have the second-order temporal coherence function for $N$ RGs,
\begin{eqnarray}\label{g2-n5}
G^{(2)}(t_{1}-t_{2}) \propto \prod_{j=1}^{N} [1+ \text{sinc}^2 \frac{\Delta\omega_j(t_{1}-t_{2})}{2}],
\end{eqnarray}
where $\Delta\omega_j$ is the frequency bandwidth of pseudothermal light scattered by RG$_j$. The time difference between two photons at RG$_j$ is equal to $t_{1}-t_{2}$ due to point light sources are assumed in the calculations \cite{zhou-2017}. Equation (\ref{g2-n5}) becomes the common second-order coherence function of thermal light when $N$ equals 1 \cite{loudon-book,shih-book}. Two-photon superbunching is expected when $N$ is larger than 1. For instance, the degree of the second-order coherence \cite{loudon-book}, $g^{(2)}(0)$, equals 4 for two RGs. When there are $N$ RGs,  $g^{(2)}(0)$ equals $2^N$.

\section{Experiment}\label{experiment}
\begin{figure}[htb]
    \centering
    \includegraphics[width=88mm]{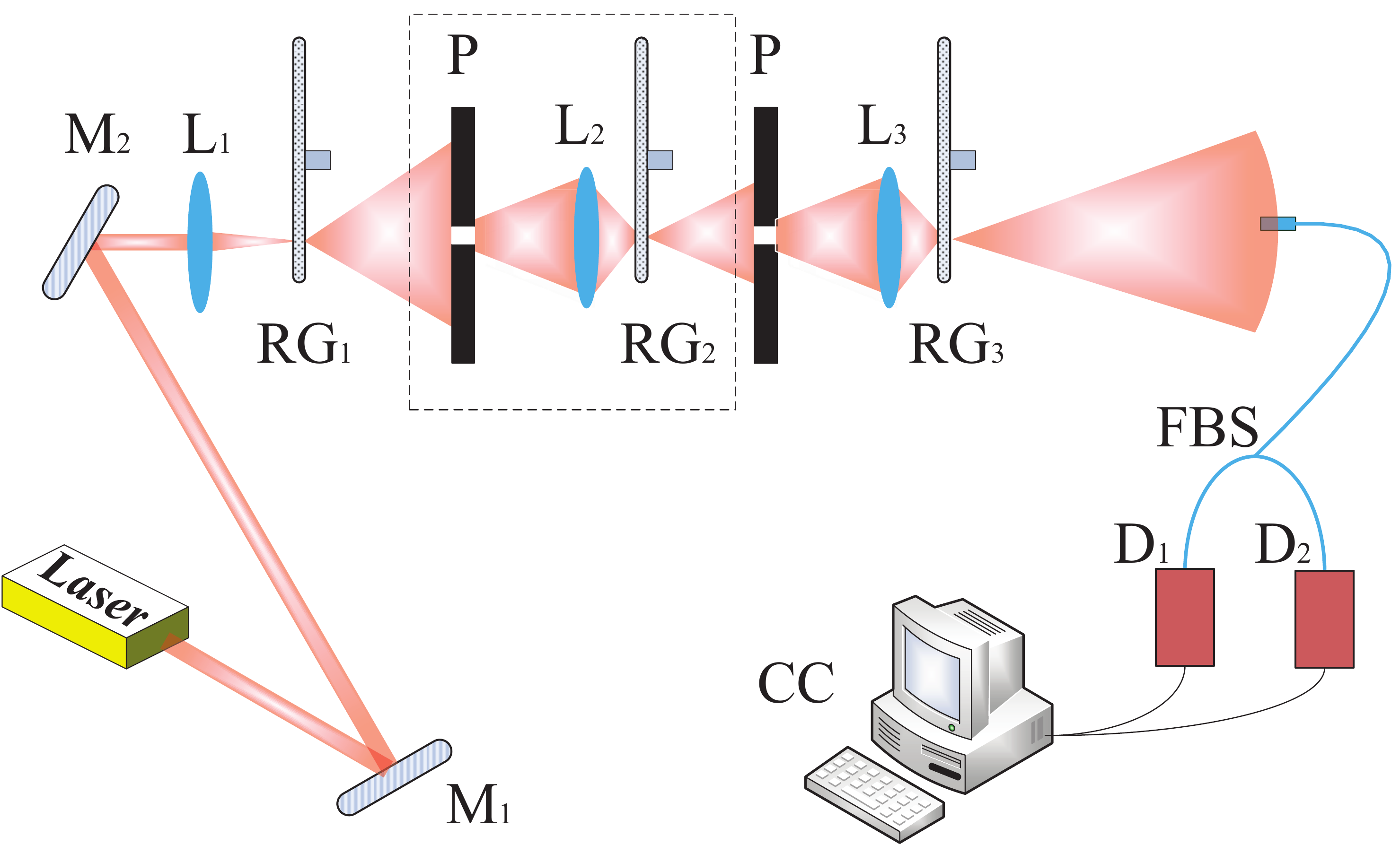}
    \caption{Experimental setup to measure two-photon superbunching of pseudothermal light. Laser: single-mode continuous-wave laser. M: mirror. L: lens. RG: rotating ground-glass. P: pinhole. FBS: $50:50$ non-polarized fiber beam splitter. D: single-photon detector. CC: two-photon coincidence count detection system. See text for details.}
    \label{3}
\end{figure}

The experimental setup to observe two-photon superbunching is shown in Fig. \ref{3}.  The employed laser is a linearly polarized single-mode continuous-wave laser with central wavelength at 780 nm and frequency bandwidth of 200 kHz. M$_1$ and M$_2$ are two mirrors. A lens with 50 mm focus length (L$_1$) is employed to focus the laser light onto RG$_1$. After propagating some distance, the scattered light is filtered by a pinhole (P). The filtered light is then focused by another lens (L$_2$) before it is incident to another rotating ground-glass (RG$_2$). The distance between L$_1$ and RG$_1$ is 50 mm. The diameter of the pinhole is less than the transverse coherence length of pseudothermal light generated by RG$_1$ in the pinhole plane. The filtered light is within one coherence area \cite{loudon-book,zhou-2017}. The focus length of L$_2$ is 25 mm and is employed to focus the scattered light onto RG$_2$. The distance between L$_2$ and RG$_2$ is 28 mm, which is determined by minimizing the size of light spot on RG$_2$. The experimental elements within the square dot line can be repeated many times as long as there is enough light intensity left. In our experiments, we measured the second-order coherence functions for one, two, and three RGs in the scheme shown in Fig. \ref{3}. FBS is a $50:50$ non-polarized fiber beam splitter. The diameter of the fiber in FBS is 5 $\mu$m, which is less than the transverse coherence length of pseudothermal light generated by RG$_3$ in the plane of the collector of FBS. D$_1$ and D$_2$ are two single-photon detectors. CC is two-photon coincidence count detection system. The single-photon dark counts for two detectors are around 100 c/s. The single-photon counts of two detectors are about 5000 c/s during the whole measurement. If the laser and all the elements before the collector of FBS are treated as a light source, the experimental setup in Fig. \ref{3} is a standard HBT interferometer measuring the second-order temporal coherence function \cite{hbt,hbt2}.

\begin{figure}[htb]
     \centering
    \includegraphics[width=80mm]{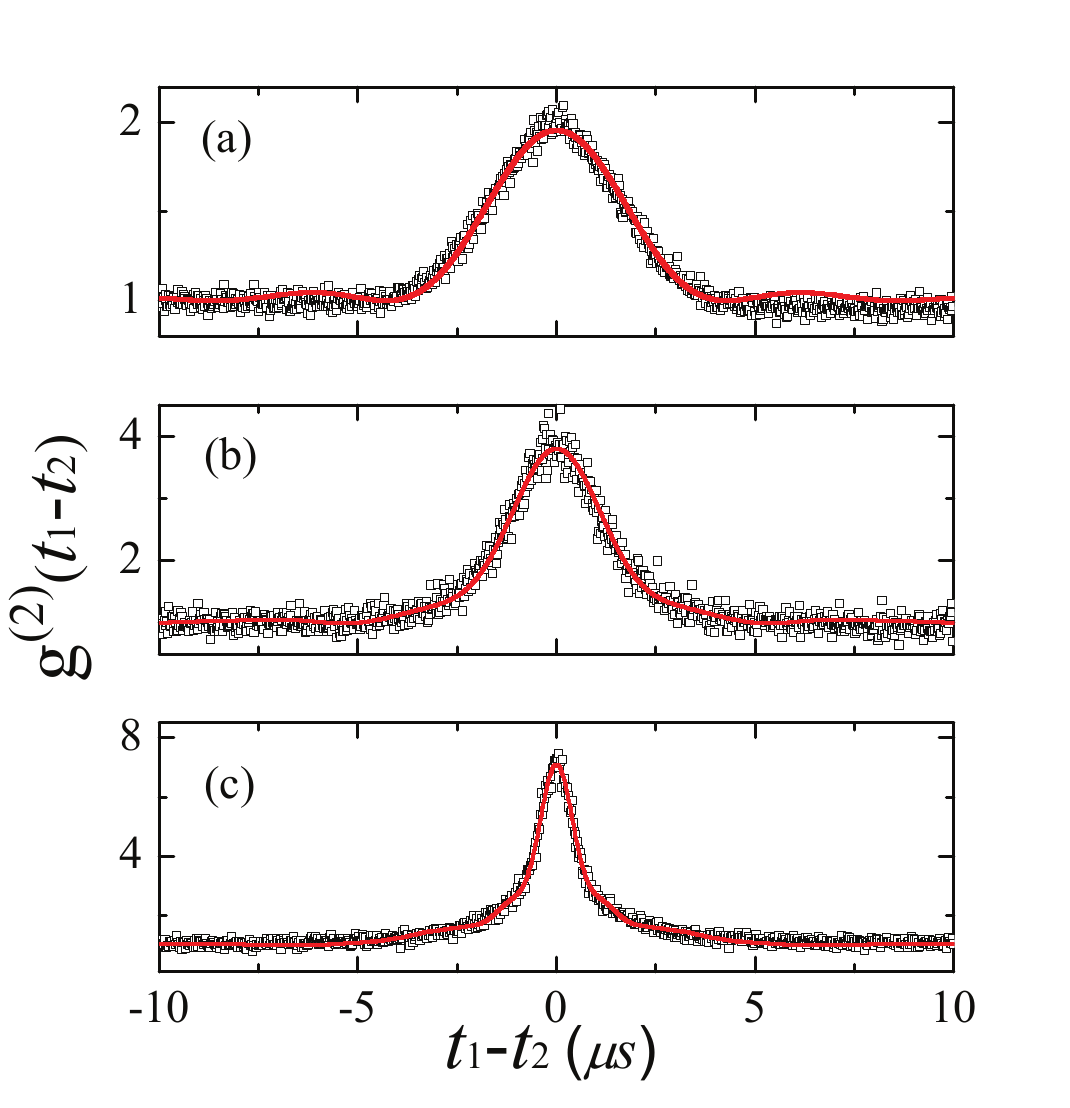}
   \caption{Measured second-order temporal coherence functions. (a), (b), and (c) are the results for one, two, and three RGs, respectively.  $g^{(2)}(t_1-t_2)$ is the normalized second-order coherence function. $t_1-t_2$ is the time difference between the two photon detection events within a two-photon coincidence count. The squares are measured data points. The red curves are theoretically fittings. Two-photon bunching is observed in (a). Two-photon superbunching is observed in (b) and (c).}
   \label{4}
\end{figure}

Figure \ref{4} shows the measured second-order temporal coherence functions for different number of RGs in the superbunching pseudothermal light scheme.  $g^{(2)}(t_1-t_2)$ is the normalized second-order coherence function and $t_1-t_2$ is the time difference between the two photon detection events within a two-photon coincidence count. The squares are measured data points, which are normalized according to the background coincidence counts. The red curves in Figs. \ref{4}(a)-\ref{4}(c) are theoretical fitting by employing Eq. (\ref{g2-n5}) for $N$ equals 1, 2, and 3, respectively.  There is only one RG for the results in Fig. \ref{4}(a), which is measured by removing RG$_2$, RG$_3$, and other related optical elements in the experimental setup shown in Fig. \ref{3}. It is a typical result for the second-order temporal coherence function of pseudothermal light \cite{martienssen}. The measured $g^{(2)}(0)$ equals $1.96\pm0.01$, which is close to the theoretical value, 2 \cite{loudon-book,martienssen}. The measured coherence time is $4.29\pm0.06$ $\mu$s in Fig. \ref{4}(a).

Figure \ref{4}(b) is the measured second-order temporal coherence function when there are two RGs in the experimental setup shown in Fig. \ref{3}. The measured $g^{(2)}(0)$ equals $3.80\pm0.04$, which is larger than 2. The measured coherence time of pseudothermal light scattered by two RGs is $2.70\pm0.07$ and  $5.26\pm0.14$  $\mu$s, respectively. The red curve is theoretical fitting of the measured data points by setting $N=2$ in Eq. (\ref{g2-n5}). Figure \ref{4}(c) is the measured second-order coherence function of pseudothermal light when there are three RGs. The measured $g^{(2)}(0)$ equals $7.10\pm0.07$. The measured coherence time scattered by three RGs is $0.96\pm0.03$, $2.24\pm0.05$, and $6.64\pm 0.15$ $\mu$s, respectively. The measured results in Figs. \ref{4}(b) and \ref{4}(c) indicate that two-photon superbunching is observed when there are more than one RGs in the scheme shown in Fig. \ref{3}.

\begin{figure}[htb]
     \centering
    \includegraphics[width=80mm]{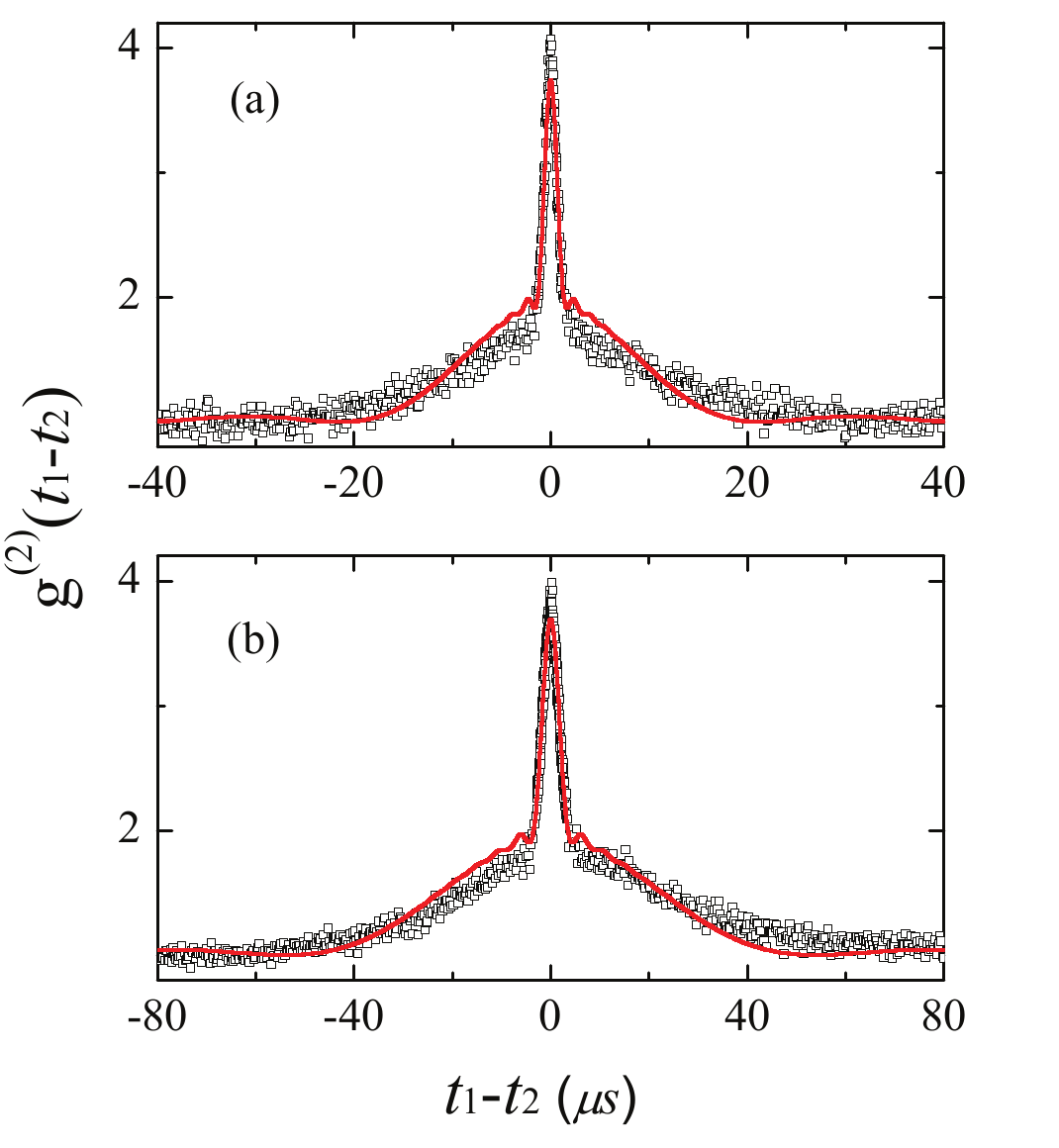}
   \caption{Measured second-order temporal coherence functions for two RGs with different rotation speeds.  (a) is measured when RG$_1$ and RG$_2$ are rotating at 30 and 3 Hz, respectively.  (b) is measured when RG$_1$ and RG$_2$ are rotating at 3 and 30 Hz, respectively. The squares are measured data points and red curves are theoretically fittings. The meanings of the symbols are similar as the ones in Fig. \ref{4}.}
   \label{5}
\end{figure}

In order to study the dependence of second-order coherence function on the coherence time of pseudothermal light scattered by each RG, we measured the second-order coherence functions when two RGs rotate at very different speeds. Figure \ref{5}(a) shows the second-order temporal coherence function when  RG$_1$ and RG$_2$ are rotating at 30 and 3 Hz, respectively. The measured $g^{(2)}(0)$ equals $3.74\pm0.05$. The measured coherence time of pseudothermal light scattered by RG$_1$ and RG$_2$ is $1.62\pm0.05$ and $21.44\pm0.56$ $\mu$s, respectively. The measured second-order coherence function in Fig. \ref{5}(a) is a product of two peaks with very different widths. The results in Fig. \ref{5}(b) are similar as the ones in Fig. \ref{5}(a) except the second-order coherence function is measured when  RG$_1$ and RG$_2$ are rotating at 3 and 30 Hz, respectively. The measured $g^{(2)}(0)$ equals $3.69\pm0.03$. The coherence time of pseudothermal light scattered by RG$_1$ and RG$_2$ is $53.97\pm0.97$ and $4.32\pm0.07$  $\mu$s, respectively. The reasons why the coherence time in Figs. \ref{5}(a) and \ref{5}(b) is different when the speeds of rotation are the same are as follows. One reason is the sizes of  light spot on these two RGs are different. Another reason is the distances between the centers of light spot and RG are not equal for two RGs. These two factors will influence the coherence time of pseudothermal light \cite{martienssen,goodman-book}.

\section{Discussion}\label{discussion}

In the last two sections, we have theoretically and experimentally proved that two-photon superbunching can be observed in our scheme. In this section, we will discuss why two-photon superbunching can be observed  and generalize it to three- and multi-photon superbunching.

From quantum optical coherence point of view, the key to have two-photon superbunching is to have more than two alternatives to trigger a two-photon coincidence count in a HBT interferometer \cite{hong,zhou-2017}.  As stated in Sec. \ref{theory}, the premise to have  Eq.  (\ref{g2-n4}) is that all the different alternatives to trigger a two-photon coincidence count are in principle indistinguishable.  Probability amplitudes are summed to calculate the second-order coherence function in this case and there is two-photon interference \cite{feynman-path}. The constructive two-photon interference is the reason why two-photon bunching \cite{fano,feynman-qed,scully-book,shih-2005} and two-photon superbunching \cite{hong,zhou-2017} can be observed. The pinhole after every RG is employed to ensure that photons passing through the pinhole are indistinguishable.

\begin{figure}[htb]
     \centering
    \includegraphics[width=55mm]{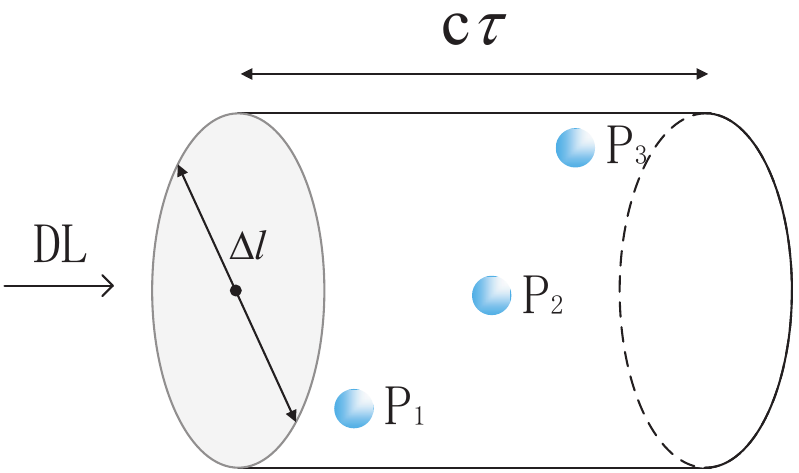}
   \caption{Coherence volume of thermal light.  DL is the direction of light propagation. $\Delta l$ is the transverse coherence length. $\tau$ is the coherence time. The coherence volume of thermal light equals the product of transverse coherence area and longitudinal coherence length. Transverse coherence area equals $\pi (\Delta l/2)^2$. Longitudinal coherence length equals $c\tau$, where $c$ is the velocity of light in the vacuum. P$_1$, P$_2$, and P$_3$ are three photons at different places within a coherence volume, which are in principle indistinguishable.}
   \label{6}
\end{figure}

Martienssen and Spiller had proved that photons within the same coherence volume are in principle indistinguishable based on the Uncertainty Principle \cite{martienssen}.  Figure \ref{6} shows the coherence volume of thermal light, which equals the product of transverse coherence area and longitudinal coherence length \cite{mandel-book}. DL is the propagation direction of thermal light. If a circular thermal light source is employed, the transverse coherence length of thermal light, $\Delta l$, equals $\lambda L/D$, where $\lambda$ is the wavelength of light, $L$ is the distance between source and detection planes, and $D$ is the diameter of thermal light source. $\tau$ is the coherence time of thermal light, which is determined by the frequency bandwidth of thermal light. P$_1$, P$_2$, and P$_3$ are three photons within the same coherence volume, which are indistinguishable.  The diameter of the employed pinhole in our experiment is less than the transverse coherence length of pseudothermal light generated by RG before the pinhole, which guarantees that the photons passing through the pinhole are within the same coherence area. The two-photon coincidence time window, which is controlled by the two-photon coincidence count detection system (CC), is much shorter than the coherence time of pseudothermal thermal light. These two conditions ensure that all the detected photons are within the same coherence volume, which guarantees that all the different alternatives to trigger a two-photon coincidence count in the HBT interferometer are indistinguishable.

The observed two-photon superbunching in our scheme can also be understood in classical theory \cite{glauber2,sudarshan}. We will follow the method in Goodman's book to calculate the degree of the third- and higher-order coherence of superbunching pseudothermal light \cite{zhou-2017,goodman-book}. The intensity of light scattered by one RG follows negative exponential distribution \cite{goodman-book},
\begin{eqnarray}\label{g2c1}
P_{I|x}(I|x)=\frac{1}{x}\text{exp}(-\frac{I}{x}),
\end{eqnarray}
where $x$ is the average intensity of light in the detection plane and is proportional to the intensity of the incident light. In the original pseudothermal light source introduced by Martienssen and Spiller, the incident light is single-mode continuous-wave laser light, in which $x$ is a constant \cite{martienssen}. If the incident light is filtered by a pinhole as the one in our scheme, the intensity, $x$, obeys the negative exponential distribution, too. The density distribution of the light intensity after RG$_2$ is
\begin{eqnarray}\label{g2c2}
P_{I}(I)=\int_{0}^{\infty}\frac{1}{x}\text{exp}(-\frac{I}{x})\cdot \frac{1}{\langle I \rangle }\text{exp}(-\frac{x}{\langle I \rangle })dx,
\end{eqnarray}
where $\langle I \rangle $ is the average intensity of the scattered light after RG$_2$. The $q$th moments of the intensity are \cite{goodman-book}
\begin{eqnarray}\label{g2cq}
\langle I^q \rangle =\int_{0}^{\infty}I^q P_I(I)dI=\langle I \rangle^q (q!)^2.
\end{eqnarray}

With the result in Eq. (\ref{g2cq}), it is ready to calculate the degree of $n$th-order coherence of pseudothermal light after two RGs. The normalized $n$th-order coherence function is defined as \cite{mandel-book}
\begin{eqnarray}\label{g2-def}
g^{(n)}(\vec{r}_1,t_1;...;\vec{r}_n,t_n)=\frac{\langle I(\vec{r}_1,t_1)...I(\vec{r}_n,t_n) \rangle}{\langle I(\vec{r}_1,t_1)\rangle ... \langle I(\vec{r}_n,t_n) \rangle},
\end{eqnarray}
where $I(\vec{r}_j,t_j)$ is the light intensity at space-time coordinates $(\vec{r}_j,t_j)$  ($j=1$, 2, ..., and $n$). When all the detectors are at the same space-time coordinates, the degree of the $n$th-order coherence is
\begin{eqnarray}\label{g2-def0}
g^{(n)}(0)=\frac{\langle I^n \rangle }{\langle I \rangle^n }.
\end{eqnarray}
Substituting Eq. (\ref{g2cq}) into Eq. (\ref{g2-def0}), it is easy to calculate the degree of $n$th-order coherence, $g^{(n)}(0)$, equals $(n!)^2$. For instance, $g^{(2)}(0)$ equals 4, which is consistent with Eq. (\ref{g2-n5}) calculated in quantum theory. The degree of third-order coherence equals 36, which is larger than the one of thermal light, 6 \cite{liu-2009}. Hence three-photon superbunching can also be observed. The degree of $n$th-order coherence of thermal light equals $n!$ \cite{liu-2009}, which means that $n$-photon superbunching is expected in the two-RG scheme for $n$ greater than 1.

The same method can be employed to calculate the scheme for more than two RGs. The $q$ moments of the intensity after $m$ ($m=2$, 3, 4, and 5) RGs is \cite{zhou-2017}
\begin{eqnarray}\label{g2cqm}
\langle I^q \rangle =\int_{0}^{\infty}I^q P_I(I)dI=\langle I \rangle^q (q!)^m.
\end{eqnarray}
The degree of $n$th-order coherence of pseudothermal light after $m$ RGs in our scheme equals
\begin{eqnarray}
g^{(n)}(0)=(n!)^m.
\end{eqnarray}
The degree of coherence can be very large if more RGs are employed. For instance, the degree of second-order coherence of pseudothermal light with five RGs equals $2^5$ (=32), which is much larger than 2. The degree of third-order coherence of pseudothermal light with five RGs equals $6^5$ (=7776), which greatly exceeds the one of thermal light, 6.

\section{Conclusion}\label{conclusion}

In conclusion, we have proved that the degree of coherence of pseudothermal light can be tuned by modulating its phase and intensity via rotating ground glasses, lenses, pinholes, \textit{etc.}. Two-, three- and multi-photon superbunching can be observed in the proposed scheme. $g^{(2)}(0)$ equals $7.10\pm0.07$ is experimentally observed  by employing three rotating ground glasses. The degree of second-order coherence can be increased to $2^N$ if $N$ rotating ground glasses are employed. This type of superbunching pseudothermal light is important for improving the visibility of temporal ghost imaging with thermal light \cite{gi-t1}. If two-photon superbunching can be realized in the spatial domain by analogy of the one in temporal domain in our scheme, the light will be of great importance to improve the quality of ghost imaging with thermal light \cite{shih-book}.

Another interesting topic worthy of noticing is that the difference between the two-photon superbunching in linear and nonlinear systems. The ratio between the peak and the background of the second-order coherence function of superbunching pseudothermal light can be as large as the one of entangled photon pairs. However, the observed two-photon superbunching in our scheme is classical, while two-photon superbunching in entangled photon pairs is nonclassical \cite{alley,aspect-1981}. The discussion about the difference between the two-photon superbunching in our scheme and the one in entangled photon pairs will be helpful to understand the physics of two-photon superbunching. Unlike it is difficult to have three- and multi-photon superbunching with nonlinear system, it is straightforward to generate three- and multi-photon superbunching in our scheme, which is helpful to study the third- and higher-order coherence of light.

\section*{Funding Information}
National Natural Science Foundation of China (NSFC) (Grant No.11404255); National Basic Research Program of China (973 Program) (Grant No.2015CB654602); Doctoral Fund of Ministry of Education of China (Grant No.20130201120013); 111 Project of China (Grant No.B14040); Fundamental Research Funds for the Central Universities.

\end{document}